\documentclass{fizik}
\usepackage{hyperref}
\hypersetup{colorlinks=true,urlcolor=black,citecolor=blue}
\usepackage{amsfonts,amssymb,amsmath,graphicx,booktabs,multirow}
\usepackage[utf8]{inputenc}
\usepackage[english]{babel}
\usepackage[dvipsnames]{xcolor}
\usepackage{epstopdf}
\usepackage{mathtools}
\usepackage{url}
\usepackage{placeins}

\yil{}\vol{}\fpage{}\lpage{}\doi{}

\title{Statistical validation of template-based parameter extraction from toy thrust distributions for future FCC-ee studies}

\author[ILGIN]{
\textbf{Fatih ILGIN\thanks{fatih.ilgin@ozyegin.edu.tr}}\\
Özyeğin University, Istanbul, Türkiye\\
\url{https://orcid.org/0000-0002-7201-8096}
}

\newcommand{\bc}{\begin{center}}
\newcommand{\ec}{\end{center}}

\renewcommand{\phi}{\varphi}

\begin{document}
\maketitle

\begin{abstract}
A statistically reliable parameter-extraction procedure should be validated independently of the physical and detector models to which it will eventually be applied. A template-based inference framework is developed and tested using controlled toy thrust distributions as a methodological benchmark for future FCC-ee event-shape studies. The model contains two-jet-like and three-jet-like event components whose relative contribution is governed by a continuous control parameter. Three extraction strategies are compared: a discrete Pearson-type $\chi^2$ distance, a discrete multinomial negative-log-likelihood estimator, and a likelihood estimator based on linearly interpolated templates. Their performance is quantified with 10,000 pseudo-experiments for samples containing 200--2000 events. Off-grid recovery, histogram-binning variations, template-statistics variations, and a model-mismatch test based on modified momentum smearing are also studied. The multinomial likelihood consistently outperforms the simple uncorrelated Pearson-type baseline used here. For an input parameter of 0.30, the correct-template fraction increases from $0.467\pm0.005$ for 200 events to $0.934\pm0.002$ for 2000 events. Interpolation reduces off-grid discretization effects, with a maximum absolute mean bias of approximately 0.003 in the tested configuration. Model mismatch nevertheless produces measurable biases, demonstrating that nominal closure alone is insufficient to establish robustness. This study does not determine the physical strong coupling constant. It provides a controlled framework for comparing estimators, quantifying statistical resolution, and diagnosing model dependence before application to realistic parton-shower simulations, hadronization models, detector effects, and FCC-ee data.
\end{abstract}

\section{Introduction}
\label{sec:intro}
The strong coupling constant $\alpha_s$ is a fundamental parameter of quantum chromodynamics (QCD) and enters precision predictions for a broad range of Standard Model processes. Its uncertainty propagates into calculations of hadronic electroweak, Higgs, and top-quark observables, so continued improvement of $\alpha_s$ determinations remains an important objective of precision physics \cite{denteria2024,pdg2024,alphas2015}.

Electron--positron collisions provide a clean environment for QCD because the initial state is colourless and the centre-of-mass energy is precisely defined. The FCC-ee programme is expected to produce very large samples of hadronic events at several collision energies, enabling precision studies of event shapes, jet rates, parton showers, fragmentation, and non-perturbative dynamics \cite{fccopp,fcc-ee,monni2021,fccqcd2025}. Recent phenomenological work has already studied thrust and the $C$-parameter at 91.2, 160, 240, and 365 GeV with PYTHIA, including initial-state radiation, electroweak backgrounds, and fits to perturbative QCD predictions \cite{mathew2026}.

Event-shape observables characterize the global geometry of hadronic final states and have played a central role in tests of QCD and determinations of $\alpha_s$ in electron--positron annihilation \cite{brandt1964,farhi1977,gehrmann2007,abbate2011,kardos2022}. Thrust is particularly useful because of its simple geometrical interpretation and sensitivity to additional radiation. Precision analyses, however, are affected by perturbative truncation, logarithmic corrections, hadronization, fit-range choices, and correlations between $\alpha_s$ and non-perturbative parameters. Soft-drop studies show that grooming can reduce hadronization sensitivity and improve fit stability, while leaving non-trivial dependence on observable definition and modelling choices \cite{baron2018,marzani2019}.

These complications make it difficult to determine whether a bias in a realistic analysis originates from the statistical estimator, template construction, event generator, hadronization, detector response, or theory model. Realistic FCC-ee event-shape analyses combine parton-shower evolution, hadronization, initial-state radiation, electroweak backgrounds, detector response, and theoretical uncertainties. Introducing these ingredients simultaneously can obscure whether an observed extraction bias is statistical or physical in origin. The deliberately minimal construction used here is therefore defined as a zeroth-order validation benchmark: it isolates estimator bias, finite-template fluctuations, grid discretization, and controlled model mismatch before realistic QCD complexity is introduced. This is not the only possible development strategy, but it provides a transparent diagnostic layer that can later be reused with generator-level and detector-level templates. Closure testing implements this strategy by generating pseudo-data from a known input, applying the full extraction procedure, and comparing the result with the known truth. Such tests are widely used to diagnose methodological bias and statistical faithfulness in inverse problems and global fitting frameworks \cite{deldebbio2022,deans2014}.

The present work develops a toy-model benchmark for template-based inference from thrust distributions. It does not reproduce realistic QCD radiation and does not determine physical $\alpha_s$. The question addressed is narrower: how reliably can a parameter controlling the mixture of two-jet-like and three-jet-like topologies be recovered from finite pseudo-data samples? The study compares a Pearson-type distance with multinomial likelihood estimators, quantifies bias and root-mean-square error (RMSE), examines off-grid recovery, and performs binning, template-statistics, and model-mismatch tests. The resulting workflow is intended as a reusable validation layer for future generator-level and detector-level studies. The deliberate simplicity of the toy model is therefore a feature of the validation design rather than an attempt to approximate the complete FCC-ee environment. Its purpose is to test the statistical extraction layer in isolation before physical $\alpha_s$ variations, parton showers, hadronization, and detector effects are introduced.

\section{Methodology}
\label{sec:method}
\begin{figure}[htbp]
\centering
\includegraphics[width=0.72\textwidth]{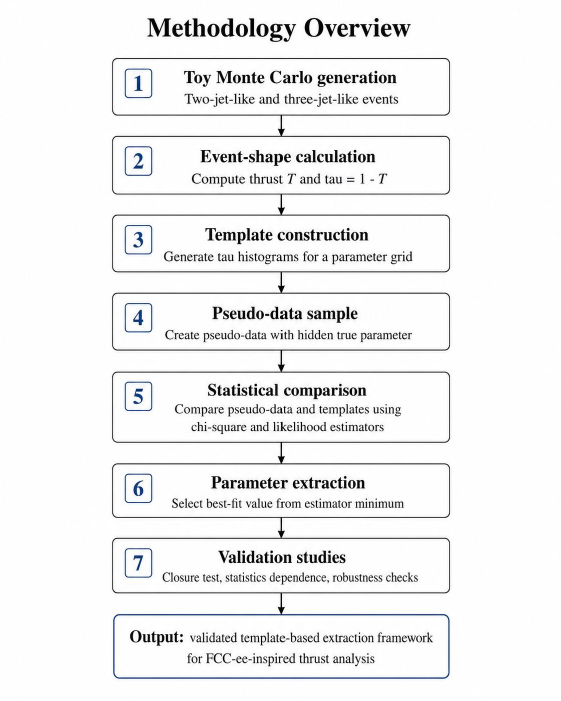}
\caption{Schematic overview of the validation workflow used in this study. Toy events are generated, thrust and $\tau=1-T$ are computed, template and pseudo-data histograms are constructed, candidate estimators are evaluated, and the extraction is validated through closure, robustness, and model-mismatch tests.}
\label{fig:workflow}
\end{figure}

This section defines the toy model, observable, template construction, statistical estimators, and validation tests used throughout the study. A schematic overview of the full workflow is shown in Figure~\ref{fig:workflow}.

\FloatBarrier

\subsection{Toy mixture model}
Each event is generated as either a two-jet-like or a three-jet-like topology. The two classes have fixed momentum-smearing models, while a continuous control parameter $p$ determines their mixture:
\begin{equation}
P(\tau\mid p)=(1-p)f_2(\tau)+pf_3(\tau),
\label{eq:mixture}
\end{equation}
where $f_2$ and $f_3$ are the normalized component densities. The discrete template grid is $p=0.10,0.15,\ldots,0.50$. The parameter is a proxy for increasing radiation activity and is not identified with physical $\alpha_s$.

\subsection{Thrust calculation}
For each event, thrust is calculated from the generated final-state momenta as
\begin{equation}
T=\max_{\hat{n}}\frac{\sum_i |\vec{p}_i\cdot\hat{n}|}{\sum_i |\vec{p}_i|},
\label{eq:thrust}
\end{equation}
\noindent and the analysed observable is $\tau=1-T$. The unsmeared two- and three-particle reference configurations satisfy global momentum conservation in the transverse plane. After independent Gaussian momentum smearing, exact event-by-event momentum balance is not reimposed. For each generated event, the thrust axis is determined by evaluating the thrust functional along the unit directions of all final-state particle momenta. This prescription is sufficient for the restricted two- and three-particle toy configurations but is not a general-purpose high-multiplicity thrust algorithm.

\subsection{Template construction}
Each baseline template contains 5000 events. The range $0\leq\tau\leq0.4$ is divided into 20 bins; 10-bin and 40-bin alternatives are tested. A pseudocount $\lambda=0.5$ stabilizes empty bins:
\begin{equation}
T_i(p)=\frac{n_i(p)+\lambda}{\sum_{j=1}^{N_{\rm bins}}[n_j(p)+\lambda]}.
\label{eq:template}
\end{equation}

\subsection{Estimators}
The baseline Pearson-type shape distance is
\begin{equation}
\chi^2(p)=\sum_{i=1}^{N_{\rm bins}}\frac{[D_i-T_i(p)]^2}{T_i(p)},
\label{eq:chi2}
\end{equation}
where $D_i$ is the normalized pseudo-data content. Equation~(\ref{eq:chi2}) is intentionally retained as a simple, uncorrelated baseline rather than as an optimal $\chi^2$ estimator. Unit-area normalization induces correlations among bins, and finite template uncertainty is not included. The comparison made below must therefore not be interpreted as establishing the superiority of likelihood methods over every covariance-aware $\chi^2$ construction. Instead, it quantifies the performance loss associated with applying this commonly accessible normalized-shape distance in the present controlled problem.

For a fixed pseudo-data event count, the binned counts follow a multinomial distribution. Ignoring terms independent of $p$, the negative log-likelihood is
\begin{equation}
-2\ln\mathcal{L}(p)=-2\sum_{i=1}^{N_{\rm bins}}d_i\ln T_i(p)+C,
\label{eq:nll}
\end{equation}
where $d_i$ is the pseudo-data count in bin $i$.

To reduce grid discretization, neighbouring templates are linearly interpolated:
\begin{equation}
T_i(p)=(1-w)T_i(p_k)+wT_i(p_{k+1}),\qquad
w=\frac{p-p_k}{p_{k+1}-p_k}.
\label{eq:interp}
\end{equation}
The interpolated likelihood is evaluated on a grid with spacing 0.001. In all cases,
\begin{equation}
\hat{p}=\operatorname*{arg\,min}_{p\in\mathcal{P}}S(p),
\label{eq:estimator}
\end{equation}
where $S$ denotes the selected score.

\subsection{Validation protocol}
The baseline input is $p_{\rm true}=0.30$. Pseudo-data samples contain 200, 500, 1000, or 2000 events, and 10,000 pseudo-experiments are generated for each condition. For discrete estimators, the correct-template fraction is
\begin{equation}
\varepsilon(N)=\frac{1}{N_{\rm PE}}\sum_{r=1}^{N_{\rm PE}}\mathbb{I}(\hat{p}_r=p_{\rm true}),
\label{eq:eff}
\end{equation}
with binomial standard error $[\varepsilon(1-\varepsilon)/N_{\rm PE}]^{1/2}$. Bias and RMSE are
\begin{equation}
\operatorname{Bias}(\hat{p})=\langle\hat{p}\rangle-p_{\rm true},\qquad
\operatorname{RMSE}(\hat{p})=\sqrt{\langle(\hat{p}-p_{\rm true})^2\rangle}.
\label{eq:metrics}
\end{equation}

Off-grid inputs $p_{\rm true}=0.125,0.175,\ldots,0.475$ are tested with 10,000 pseudo-experiments each. Robustness tests vary the number of histogram bins, template sample size, and pseudo-data momentum-smearing width. In the model-mismatch test, the pseudo-data smearing width is increased by 10\% while the nominal templates are retained. All calculations use fixed documented seeds.

\section{Results}
\label{sec:results}
\subsection{Nominal closure and estimator comparison}
\begin{figure}[htbp]
\centering
\includegraphics[width=0.82\textwidth]{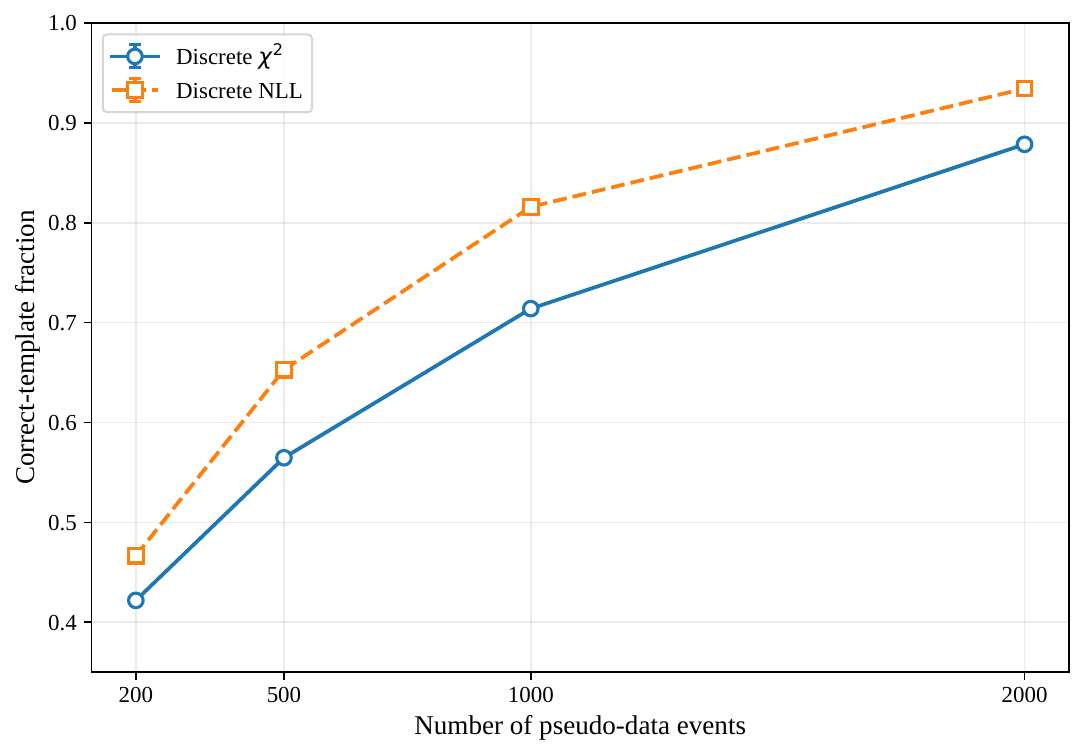}
\caption{Correct-template fraction versus pseudo-data sample size for the discrete Pearson-type $\chi^2$ and multinomial NLL estimators. Error bars are binomial standard errors from 10,000 pseudo-experiments. Distinct marker and line styles preserve readability in both colour and grayscale reproduction.}
\label{fig:correct}
\end{figure}

Figure~\ref{fig:correct} compares the correct-template fractions. Both discrete estimators improve with increasing statistics, but the multinomial likelihood performs better than the simple Pearson-type baseline at every sample size. For 200 events, the fractions are $0.422\pm0.005$ for the Pearson-type distance and $0.467\pm0.005$ for the likelihood. At 1000 events they are $0.714\pm0.005$ and $0.816\pm0.004$, and at 2000 events they reach $0.879\pm0.003$ and $0.934\pm0.002$.

\begin{table}[htbp]
\caption{Statistical performance at $p_{\rm true}=0.30$.}
\label{tab:performance}
\centering\small
\begin{tabular}{rrrrr}
\toprule
$N$ & Estimator & Bias & RMSE & Standard deviation\\
\midrule
200 & Pearson-type $\chi^2$ & 0.0119 & 0.0478 & 0.0462\\
200 & Discrete NLL & 0.0020 & 0.0401 & 0.0401\\
200 & Interpolated NLL & 0.0027 & 0.0357 & 0.0356\\
500 & Pearson-type $\chi^2$ & 0.0058 & 0.0340 & 0.0335\\
500 & Discrete NLL & 0.0015 & 0.0296 & 0.0296\\
500 & Interpolated NLL & 0.0033 & 0.0243 & 0.0241\\
1000 & Pearson-type $\chi^2$ & 0.0038 & 0.0268 & 0.0265\\
1000 & Discrete NLL & 0.0013 & 0.0215 & 0.0214\\
1000 & Interpolated NLL & 0.0047 & 0.0192 & 0.0186\\
2000 & Pearson-type $\chi^2$ & 0.0017 & 0.0174 & 0.0173\\
2000 & Discrete NLL & 0.0007 & 0.0128 & 0.0128\\
2000 & Interpolated NLL & 0.0058 & 0.0160 & 0.0149\\
\bottomrule
\end{tabular}
\end{table}

\begin{figure}[htbp]
\centering
\includegraphics[width=0.82\textwidth]{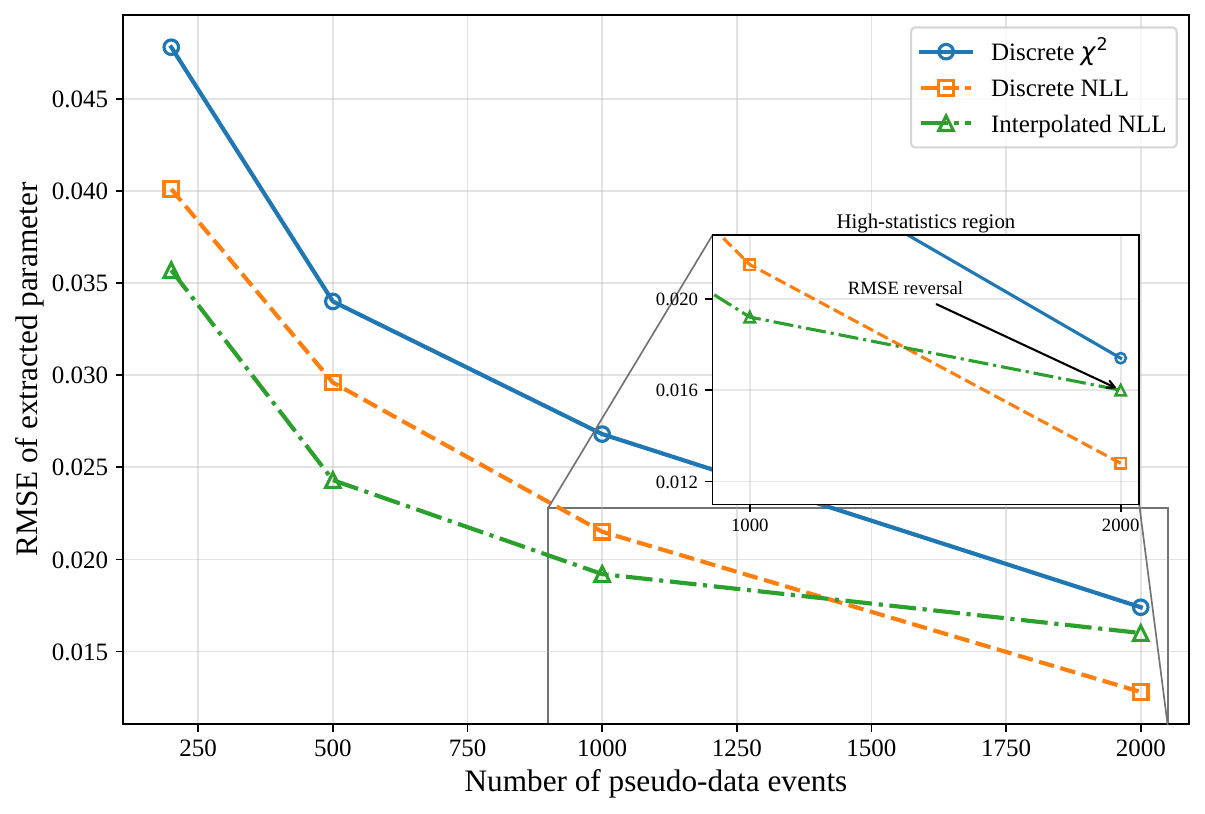}
\caption{RMSE versus pseudo-data sample size for the three estimators. The inset enlarges the high-statistics region and highlights the reversal at $N=2000$, where the interpolated NLL has a slightly larger RMSE than the discrete NLL because finite-template fluctuations limit the interpolation benefit.}
\label{fig:rmse}
\end{figure}

Figure~\ref{fig:rmse} shows that interpolation provides the smallest RMSE at 200--1000 events. The inset highlights the high-statistics region, where at 2000 events the interpolated RMSE exceeds that of the discrete likelihood, in agreement with Table~\ref{tab:performance}. This reversal is explained by the fixed size of the template library. Each template contains 5000 Monte Carlo events and therefore carries bin-level statistical fluctuations. Linear interpolation transfers not only the underlying shape evolution but also those fluctuations to intermediate parameter values. At low and moderate pseudo-data statistics, the reduction of grid quantization dominates and interpolation improves the resolution. At $N=2000$, the pseudo-data fluctuations are smaller, so finite-template structure contributes more strongly to the local likelihood slope and can shift the continuous minimum away from the true value. The discrete estimator remains constrained to the original grid point and is less sensitive to such a small continuous displacement. The deterioration is therefore interpreted as a finite-template effect, not as an intrinsic failure of interpolation.

\FloatBarrier

\subsection{Off-grid closure}
\begin{figure}[htbp]
\centering
\includegraphics[width=0.72\textwidth]{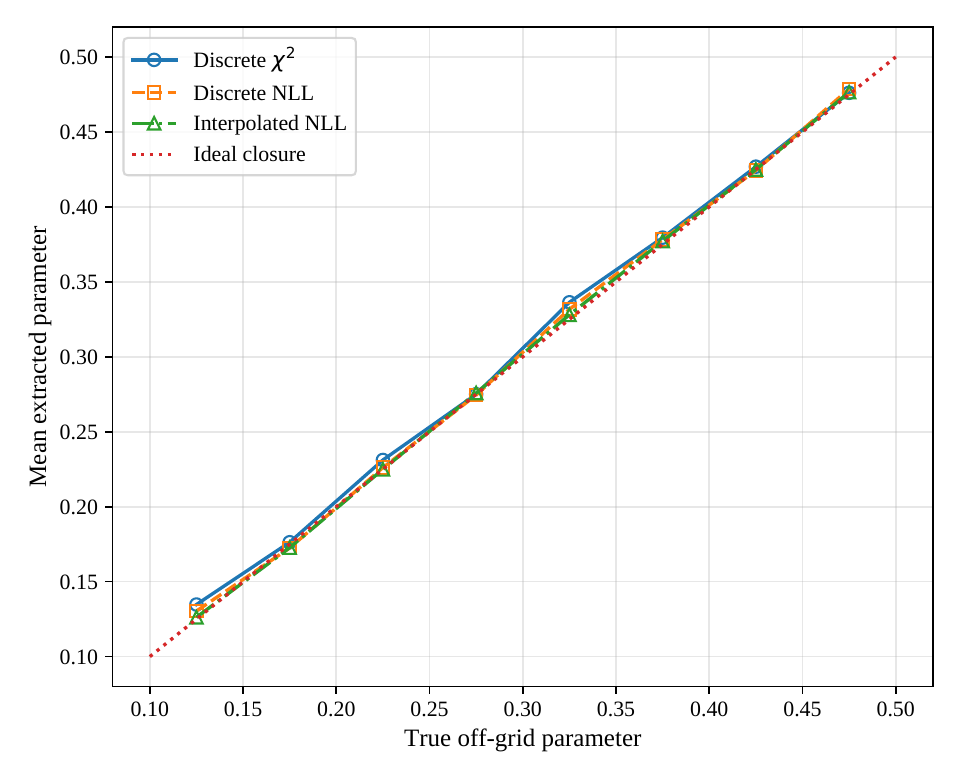}
\caption{Mean extracted parameter for off-grid inputs. The dashed line indicates ideal closure.}
\label{fig:offgrid}
\end{figure}

Figure~\ref{fig:offgrid} compares mean extracted values for inputs halfway between discrete templates. The maximum absolute mean bias is 0.0114 for the Pearson-type distance, 0.0067 for the discrete likelihood, and 0.0030 for the interpolated likelihood. The continuous method also lowers RMSE substantially, showing that interpolation mainly improves resolution and grid bias.

\FloatBarrier

\subsection{Model mismatch}
\begin{figure}[htbp]
\centering
\includegraphics[width=0.75\textwidth]{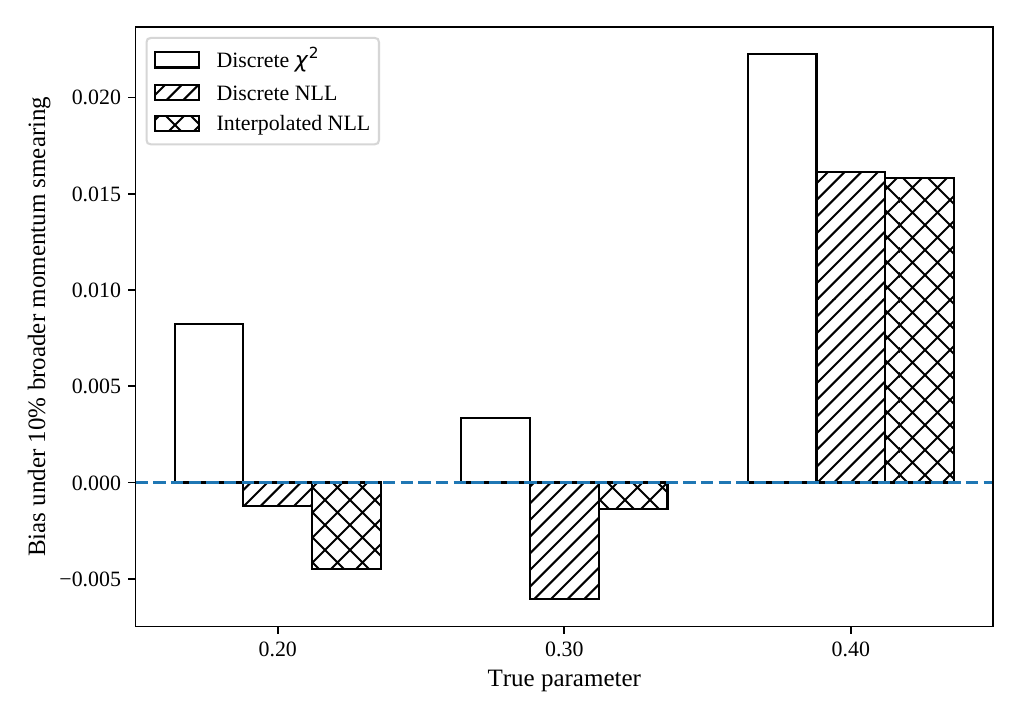}
\caption{Estimator bias when pseudo-data are generated with a momentum-smearing width 10\% larger than the nominal template value.}
\label{fig:mismatch}
\end{figure}

The model-mismatch test reveals biases that are absent under nominal closure. As shown in Figure~\ref{fig:mismatch}, at $p_{\rm true}=0.30$ the biases are 0.0034, $-0.0061$, and $-0.0014$ for the Pearson-type, discrete-likelihood, and interpolated-likelihood estimators, respectively. At $p_{\rm true}=0.40$, all methods shift upward, with biases of 0.0222, 0.0161, and 0.0158. None of the estimators is model independent.

\FloatBarrier

\subsection{Binning and template statistics}
\begin{figure}[htbp]
\centering
\includegraphics[width=0.72\textwidth]{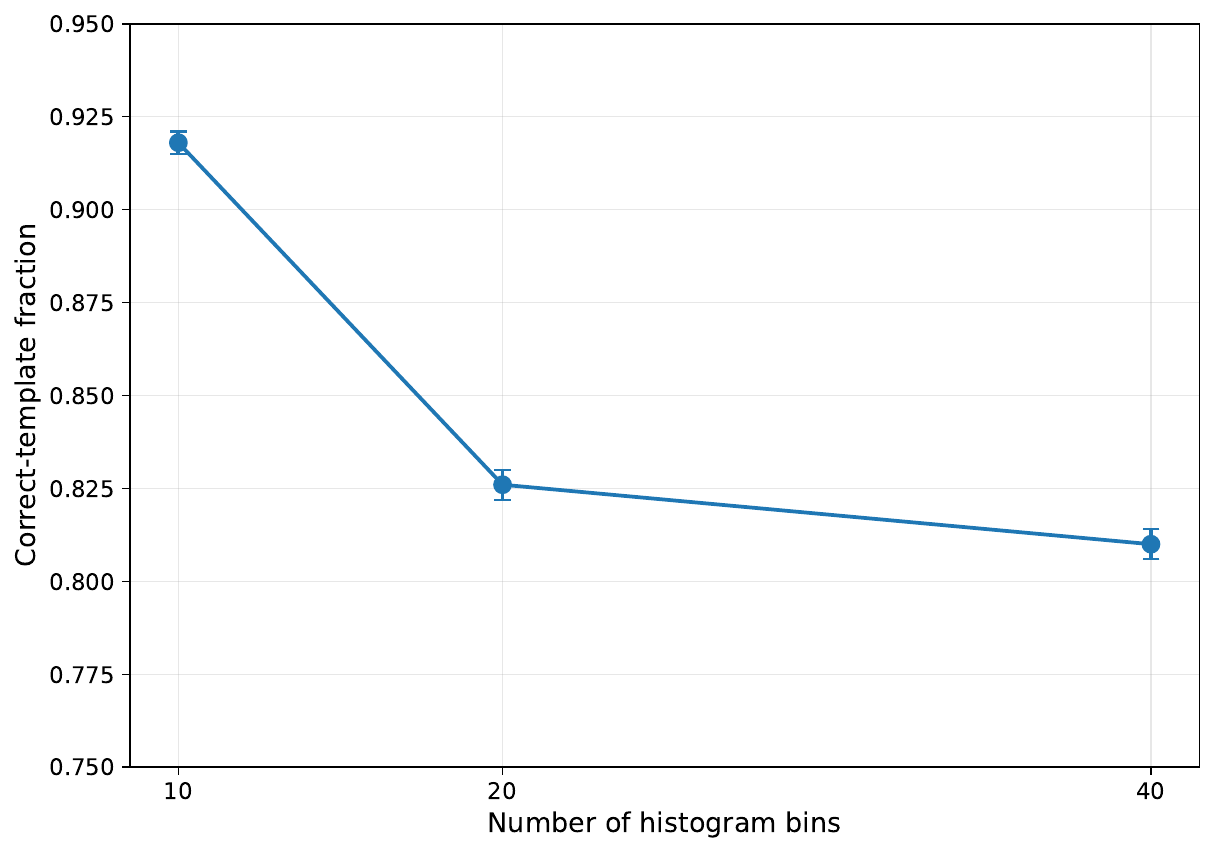}
\caption{Binning robustness of the discrete likelihood estimator at $N=1000$. Explicit circular markers indicate the three evaluated histogram-binning configurations; error bars are binomial standard errors from 10,000 pseudo-experiments.}
\label{fig:binning}
\end{figure}

Figure~\ref{fig:binning} shows that binning affects performance. The correct-template fractions for 10, 20, and 40 bins are $0.918\pm0.003$, $0.826\pm0.004$, and $0.810\pm0.004$, respectively. Finer binning does not automatically improve extraction because finite counts and template fluctuations become more important.

\begin{figure}[htbp]
\centering
\includegraphics[width=0.72\textwidth]{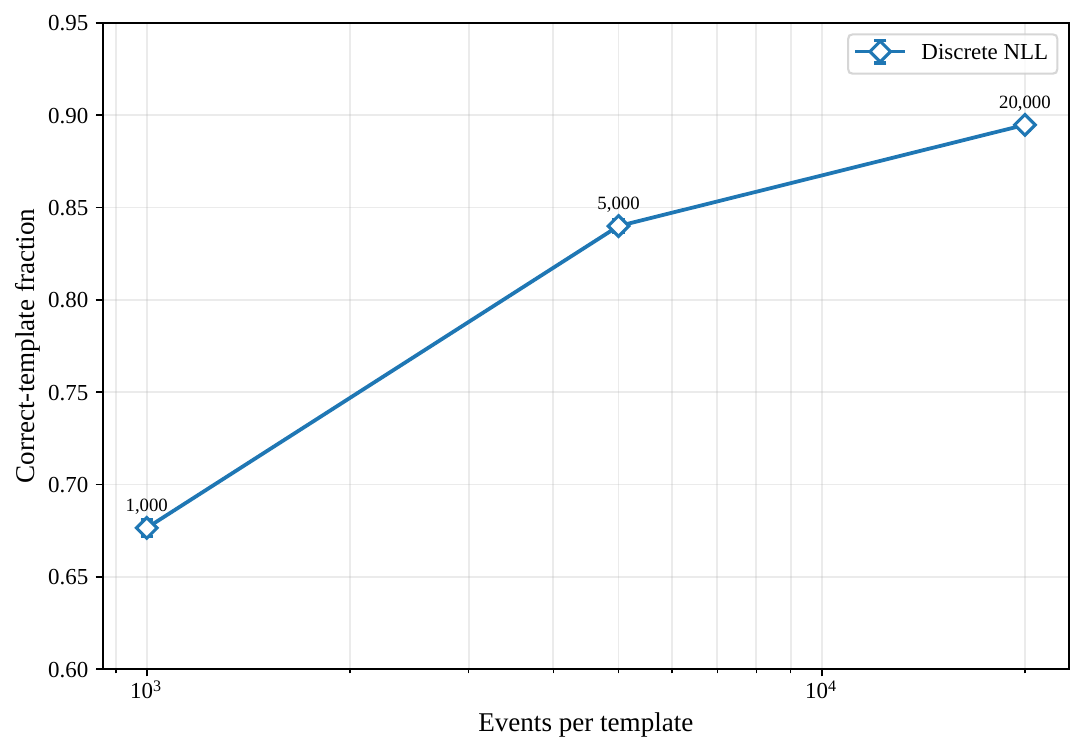}
\caption{Correct-template fraction versus the number of events used to construct each template. Explicit diamond markers indicate the three template-statistics configurations evaluated on the logarithmic horizontal axis. Error bars are binomial standard errors from 10,000 pseudo-experiments.}
\label{fig:templates}
\end{figure}

Increasing template statistics improves the result, as shown in Figure~\ref{fig:templates}. The correct-template fraction rises from $0.677\pm0.005$ with 1000 events per template to $0.895\pm0.003$ with 20,000 events. The corresponding RMSE decreases from 0.0290 to 0.0162.

\FloatBarrier

\section{Discussion}
\label{sec:discussion}
The multinomial likelihood is the statistically natural estimator for fixed-size binned pseudo-data because it uses event counts directly and treats normalized template contents as category probabilities. The Pearson-type expression used here is deliberately a simple baseline applied to normalized contents without a covariance model. The numerical comparison therefore demonstrates that this particular uncorrelated baseline is suboptimal for the present problem; it does not establish that likelihood methods outperform all possible $\chi^2$ constructions. A covariance-aware statistic, or a formulation that includes finite-template uncertainty, would constitute a distinct and potentially more competitive comparison. In a realistic experimental application, one possible alternative is a generalized statistic of the form
\begin{equation}
\chi^2_{\rm cov}(p)=[\mathbf{D}-\mathbf{T}(p)]^{T}\,\mathbf{V}^{+}\,[\mathbf{D}-\mathbf{T}(p)],
\label{eq:covchi2}
\end{equation}
where $\mathbf{V}$ contains statistical, detector, and systematic covariance contributions and $\mathbf{V}^{+}$ denotes a Moore--Penrose pseudoinverse. The pseudoinverse, or equivalently the removal of one dependent bin, is required because unit-normalized histogram contents obey a sum constraint. Finite-template uncertainties could additionally be incorporated through profiled nuisance parameters.

A discrete template grid imposes a resolution scale. Linear interpolation reduces this quantization and improves off-grid recovery, but it does not add independent physical information. Because each template is estimated from a finite event sample, neighbouring templates contain independent bin-level Monte Carlo fluctuations. Interpolation carries those fluctuations into the continuous probability model. Once pseudo-data statistics become sufficiently high, the uncertainty of the fixed template library can become comparable to the pseudo-data shape variation and distort the local likelihood slope. This mechanism explains why the interpolated RMSE at $N=2000$ is larger than the discrete-likelihood RMSE. Realistic applications should therefore propagate template uncertainty through suitably constrained nuisance parameters or larger independent template samples.

The distinction between nominal closure and robustness is central. Nominal closure verifies the analysis chain under its own generating assumptions. The smearing variation tests sensitivity to a violation of those assumptions and produces parameter-dependent biases. In future applications, momentum-smearing and detector-response uncertainties could be introduced as constrained nuisance parameters and profiled jointly with the parameter of interest, thereby allowing the fit to propagate such modelling uncertainties and potentially reduce the associated bias. This is qualitatively consistent with realistic event-shape analyses, where hadronization, fit range, radiation modelling, and detector effects can shift extracted parameters even when the statistical fit is well behaved \cite{mathew2026,baron2018,marzani2019}.

The principal limitations are the simplified two-component model, restricted low-multiplicity thrust calculation, absence of physical $\alpha_s$ variation, and use of only one prescribed model deformation. The next stage is to replace the toy mixture with independent parton-shower samples generated at several physical $\alpha_s$ values, followed by hadronization and fast detector simulation. The same closure, off-grid, binning, template-statistics, and mismatch diagnostics can then be applied without changing the statistical logic.

\section{Conclusion}
A controlled template-based inference framework has been validated through direct estimator comparison and multiple robustness tests. The multinomial negative-log-likelihood estimator consistently outperforms the simple uncorrelated Pearson-type baseline in nominal closure. Interpolated templates improve off-grid recovery, although finite template statistics limit their benefit at high pseudo-data statistics. Model mismatch produces measurable biases and demonstrates that successful self-closure is not sufficient evidence of robustness. The workflow therefore provides a quantitative benchmark for future generator-level and detector-level FCC-ee event-shape studies, while making no claim of a physical $\alpha_s$ determination.

\section*{Declaration of generative AI and AI-assisted technologies}
During the preparation of this work, the author used OpenAI ChatGPT for English translation and LaTeX preparation. The author independently executed the code, verified the numerical results and figures, reviewed the manuscript, and takes full responsibility for the work.

\section*{Data availability statement}
The complete analysis script, fixed configuration, numerical tables, and figure files are provided as ancillary material accompanying this preprint. No external experimental dataset was used.

\section*{Funding}
The author received no specific funding for this work.

\section*{Conflict of interest}
The author declares no conflict of interest.


\begin{thebibliography}{99}
\bibitem{brandt1964} S. Brandt, C. Peyrou, R. Sosnowski, and A. Wroblewski, Phys. Lett. 12, 57 (1964). doi:10.1016/0031-9163(64)91176-X.
\bibitem{farhi1977} E. Farhi, Phys. Rev. Lett. 39, 1587 (1977). doi:10.1103/PhysRevLett.39.1587.
\bibitem{gehrmann2007} A. Gehrmann-De Ridder, T. Gehrmann, E.W.N. Glover, and G. Heinrich, JHEP 12, 094 (2007). doi:10.1088/1126-6708/2007/12/094.
\bibitem{abbate2011} R. Abbate, M. Fickinger, A.H. Hoang, V. Mateu, and I.W. Stewart, Phys. Rev. D 83, 074021 (2011). doi:10.1103/PhysRevD.83.074021.
\bibitem{denteria2024} D. d'Enterria et al., J. Phys. G: Nucl. Part. Phys. 51, 090501 (2024). doi:10.1088/1361-6471/ad1a78.
\bibitem{pdg2024} S. Navas et al. (Particle Data Group), Phys. Rev. D 110, 030001 (2024). doi:10.1103/PhysRevD.110.030001.
\bibitem{fccopp} A. Abada et al. (FCC Collaboration), Eur. Phys. J. C 79, 474 (2019). doi:10.1140/epjc/s10052-019-6904-3.
\bibitem{fcc-ee} A. Abada et al. (FCC Collaboration), Eur. Phys. J. Spec. Top. 228, 261 (2019). doi:10.1140/epjst/e2019-900045-4.
\bibitem{alphas2015} D. d'Enterria and P.Z. Skands, eds., High-Precision $\alpha_s$ Measurements from LHC to FCC-ee, Proceedings of the Workshop held at CERN, Geneva, 12--13 October 2015, CERN-PH-TH-2015-299, arXiv:1512.05194 [hep-ph] (2015).
\bibitem{monni2021} P.F. Monni and G. Zanderighi, Eur. Phys. J. Plus 136, 1162 (2021). doi:10.1140/epjp/s13360-021-02105-4.
\bibitem{deldebbio2022} L. Del Debbio, T. Giani, and M. Wilson, Eur. Phys. J. C 82, 330 (2022). doi:10.1140/epjc/s10052-022-10297-x.
\bibitem{deans2014} C.S. Deans, Closure testing the NNPDF3.0 methodology, Proceedings of QCD14, pp. 15--18 (2014), arXiv:1409.4283 [hep-ph].
\bibitem{kardos2022} A. Kardos, G. Somogyi, and A. Verbytskyi, SciPost Phys. Proc. 10, 014 (2022). doi:10.21468/SciPostPhysProc.10.014.
\bibitem{fccqcd2025} D. d'Enterria, P.F. Monni, P. Skands, and A. Verbytskyi, ``Physics case for low-$\sqrt{s}$ QCD studies at FCC-ee,'' CERN-TH-2025-064, arXiv:2503.23855 [hep-ex] (2025).
\bibitem{mathew2026} P. Mathew, R. Aggarwal, and M. Kaur, Phys. Rev. D 113, 116017 (2026). doi:10.1103/243z-g9x8.
\bibitem{baron2018} J. Baron, S. Marzani, and V. Theeuwes, JHEP 08, 105 (2018). doi:10.1007/JHEP08(2018)105.
\bibitem{marzani2019} S. Marzani, D. Reichelt, S. Schumann, G. Soyez, and V. Theeuwes, JHEP 11, 179 (2019). doi:10.1007/JHEP11(2019)179.
\end{thebibliography}
\end{document}